# Stepwise emergence of an intensive physical property from a single-atom to bulk

Jason N. Armstrong, Susan Z. Hua, and Harsh Deep Chopra*

*Laboratory for Quantum Devices, Materials Program, Mechanical and Aerospace Engineering Department, The State University of New York at Buffalo, Buffalo, NY 14260, USA*

**Abstract**

Intensive or size-invariant physical properties are well known to become size-dependent when the bulk material is reduced to the nanometer scale. Using silver, the present study shows a remarkable emergent characteristic of extensive properties as the size of the system is increased from a single atom to bulk – the property (strength) evolves in a staircase manner, as opposed to the intuitively assumed continuous approach to a saturating bulk value. In other words, the observed variation with size retains the inherent trait of intensive properties in the form of size-independent staircase plateaus. The steps can be explained by simple and necessary geometric configurations that atoms must assume in their initial approach to bulk. Were it not for these observations, it would have been remarkable to have reported that the observed strength in the limit of a single-atom Ag bridge approaches the theoretical limit, being 3-4 orders of magnitude higher than in bulk single crystals. The appearance of steps is acutely sensitive to minute perturbations, and their observation is enabled by pico-level resolution in measured forces and displacements in highly stable atomic sized bridges. However, suitably harnessed, results open the possibility to impart vastly different strengths to the same material based only on its size. This combined with possibility to change the position and height of the steps by mixing elements of different diameters and surface energy, opens a realistic approach to 'materials by design'. Results also provide an example of the evolutionary trace of an emergent physical property in materials.

Intensive physical properties such as melting point, modulus, strength, magnetization, heat capacity, etc. are size-independent. However, it is well known that when the system is reduced from bulk to nanometer scale, surface and confinement effects begin to dominate, and these properties diverge from their bulk values.[1-19] Such studies not only demarcate the size below which the classical description of 'bulk' properties is no longer applicable, but also provide means to investigate the evolutionary trace of emergent physical property.[20] Our recent results on mechanics of atomic-sized gold bridges having conductance in the quantum and Sharvin regime have shown a remarkable modulus enhancement as the size of the sample approaches the Fermi wavelength of the electrons.[9] These results also revealed two fundamental crossovers in deformation modes with increasing contact diameter, first, from homogeneous shear to defect mediated deformation that is in close agreement with previous predictions,[21] followed by another transition from surface to volume dominated deformation. These results build on previous experimental and theoretical studies on atomistic processes underlying the observed mechanics and transport behavior, using gold as a model system.[21-28] Experimentally, the mechanical behavior of these atomic-sized samples is studied by forming a sample between a substrate and an atomically sharp tip of the same material (say, gold). While gold has been widely studied, it is less suitable for investigating the role of surface energy γ. This is because the surface energy of gold varies significantly with the crystallographic planes; $\gamma_{(111)}$, $\gamma_{(100)}$, and $\gamma_{(110)}$ being 1283 mJ/m$^2$, 1627 mJ/m$^2$, and 1700 mJ/m$^2$, respectively, a variation of ~33%.[29] During repeated experiments, samples of random crystallographic orientations form between the tip and the substrate, leading to large variations in surface energy related effects. By contrast, the surface energy for silver has a much smaller variation with crystal orientation. The values of $\gamma_{(111)}$, $\gamma_{(100)}$, and $\gamma_{(110)}$ for silver varies by only ~6%, being 1172 mJ/m$^2$, 1200 mJ/m$^2$, and 1238 mJ/m$^2$, respectively.[29] Silver is also monovalent and isomorphous with gold (face-centered cubic). Moreover, the Ag-Ag bond length (0.2889 nm) and lattice constant (0.4085 nm) is virtually identical to that for gold (0.2884 nm and 0.4078 nm, respectively), allowing for meaningful comparisons to be made.

The Ag films (200 nm thick) were magnetron sputtered (30 W) on silicon substrates in an Ar partial pressure of 3 mtorr in a UHV chamber whose base pressure is ~$10^{-8}$-$10^{-9}$ torr. The Ag sputtering target was 99.999% pure. Simultaneously, atomic force microscope (AFM) silicon cantilever tips were also sputter coated with Ag for force-deformation measurements. During deposition the cantilevers were periodically rotated relative to the sputtering gun to enhance the uniformity of the films. Deposition conditions for gold are described previously.[9,15] A modified AFM (Ambios Q-Scope Nomad) was used for simultaneous measurements of force-deformation and conductance at room temperature in inert atmosphere. As described previously,[9] the AFM assembly consisted of a dual piezo configuration, one for the coarse and another for the fine alignment of the substrate relative to the cantilever tip. With this configuration, the minimum step size was 4 pm and the noise was ~5 pm. A range of cantilever spring constants was used (16-80 N/m) to study samples as small as a single-atom bridge; need for using cantilevers with different stiffness is discussed previously.[9] The cantilevers were precisely calibrated using reference cantilevers available from Veeco Probes (Force Calibration Cantilevers CLFC-NOBO). The photo-detector was calibrated using the well established optical deflection technique. Conductance traces for Ag and Au were recorded at a bias voltage of 100 mV and 250 mV, respectively. In all experiments, the piezo was extended or retracted at a rate of 5 nm/s. Both Ag and Au are 'noble' metals because they occur in their native form in nature; nor do they easily oxidize if suitable precautions are taken. However, silver is susceptible to sulfur, forming silver sulfide (gray patina). Therefore, the history of the deposition chamber should be free of any recently used sulfur-containing targets. If any oxide or sulfide coating is formed on the surface during the experiments, it is easily detected by observation of force in the AFM cantilever before the detection of conductance between the tip and the substrate. However, we rarely found this condition in our experiments.

Figure 1 shows typical example of simultaneously measured force and conductance traces for Ag. Hundreds of such traces were obtained by retracting the piezo, which causes an initially large diameter bridge to be progressively broken down, until it ruptures completely. Samples ranging in size from a single-atom bridge to those containing hundred of atoms can be formed, spanning the conductance regimes from quantized to semi-classical or Sharvin. The trace in Fig. 1 is qualitatively similar to that for gold.[9, 15] For any atomic configuration, say, the one labeled '$v$' in Fig. 1(a), the force increases (more negative values) until a critical force $F_{Yield}^v$ is reached, which causes a new atomic configuration '$w$' to form abruptly; the negative force in Fig. 1 denotes experiments in retraction. As shown in Fig. 1, successive atomic configurations (such as the ones labeled '$u$', '$v$', '$w$', etc.) are separated by a stepwise change in force and a stepwise change in conductance. From hundreds of such traces, $F_{Yield}^v$ for different sized atomic configurations can be measured; the simultaneously measured conductance allows the area of the constriction to be estimated using Sharvin formula,[30, 31] as described previously for gold.[9] The Sharvin formula relates the conductance $G_{Sharvin}^v$ of an atomic configuration '$v$' to its cross-section area $A$ by the relationship $G_{Sharvin}^v = (2e^2/h)(\pi A/\lambda_F^2) = G_o(\pi A/\lambda_F^2)$; here $(2e^2/h) = G_o$ is the quantum of conductance; $e$ is the quantum of charge; $h$ is Planck constant, and $\lambda_F$ is the Fermi wavelength. If necessary, for atomic bridges in the regime of quantized conductance (~1-3 atoms), the precisely known diameter of Ag and Au atoms can be used to separately determine the cross-section area.

The size dependence of $F_{Yield}^v$ for silver is plotted in Fig. 2(a). It shows that $F_{Yield}^v$ rises sharply below ~20$G_0$, corresponding to a sample cross-section area of ~1.75 nm$^2$ (or diameter of ~1.5 nm). In the limit of a single-atom bridge, $F_{Yield}^v$ approaches ~$1.4 \times 10^{10}$ Pa. This value is comparable to the theoretical strength for silver, and is about four orders of magnitude higher than the yield strength of 99.999% annealed single crystals of silver ($0.5 \times 10^6$ Pa).[32] Figure 2(b) shows a zoom-in view of $F_{Yield}^v$ versus size for Ag. The inset in Fig. 2(b) shows additional data points from different experiments, as well as standard deviation for bridges made of less than four atoms. Remarkably, Fig. 2(b) reveals a stepwise increase in strength as the system approaches the limit of a single atom; as explained later in Fig. 3, the origin of steps is purely geometrical in nature.

For comparison, the strength of Ag and Au is plotted in Fig. 2(c). It shows that the overall strength of each element increases in the limit of a single-atom bridge, and is higher for the higher surface energy Au. According to Pauling, if the atomic coordination number is reduced, the radius of the lower coordinated atoms shrinks spontaneously, leading to an associated bond strength gain or bond stiffening. However, the atomic cohesive energy is also lowered, which relates directly to the mechanically activated processes in lower atomic coordination samples. This has been verified by the so-called bond-order-length-strength correlation mechanism.[33, 34] Thus higher surface energy elements would have a comparatively higher strength than those with lower surface energy, as seen from Fig. 2(c).

Finally, the position of the steps can be qualitatively explained by the simple and necessary geometric configurations that atoms assume in their initial approach to bulk. As shown schematically in Fig. 3, the 1, 2, and 3-atom configurations are unique and therefore discrete. Once a 3-atom constriction is formed, it gives rise to three equivalent sites, where atoms 4-6 can sit. This gives rise to the first (narrow) plateau with a width of just 3 atoms; in actual data, this plateau is skewed due to multiple possibilities of completely surrounding an atom by a hexagonal ring, as shown in Fig. 3. Once the central atom takes this stable configuration, the positions '8-13' begin to be filled, marking the range of the second plateau from 7-13. The third plateau from 13-19 is associated with formation of compete hexagonal rings around *each* atom (marked 2-7 in Fig. 3) in the first outer ring. The addition of these atoms (14-19) also results in the formation of the second outer ring for a total of 19 atoms, as shown in Fig. 3. This second outer ring creates 12 new equivalent positions around it, whose completion would lead to a new step at $31G_0$ (not shown). This step would be followed by another step at $37G_0$ where each atom in the second outer ring would act as a central atom for a hexagonal ring to form around them, leading to the completion of the third outer ring (not shown). It is of interest to note that the values of 7, 19, and $37G_0$ have been seen in our recent study related to the magnitudes of discrete atomic displacements in gold during atomic reconfigurations;[9] however, the present study shows that the plateaus are too sensitive to be seen in gold due to its large variation in surface energy along different crystallographic directions. In this regard, Ag may be a unique element in the periodic table for such investigations.

Finally, it is of interest to note that a decrease in strength to zero would represent the melting of the atomic-sized constriction, as has been shown previously.[6, 8] Based on the present study, this would be expected for lower surface energy elements such as Sn (~530 mJ/m$^2$), In (~450 mJ/m$^2$) , and Pb (~3 mJ/m$^2$), and such studies are currently underway. Study of modulus enhancement of Ag is also currently underway and will be reported later. Also note that the observation of steps would be more easily seen in strength than in modulus due to the uncertainty associated with assuming the length of the samples in modulus calculations; in contrast, strength is directly obtained from the measured force.

This work was supported by the National Science Foundation, Grant Nos. DMR-0706074, and DMR-0964830, and this support is gratefully acknowledged. *Corresponding author: H.D.C.; E-mail: hchopra@buffalo.edu

**FIGURE CAPTIONS**

**FIG. 1.** Simultaneously measured force and conductance during deformation of atomic-sized bridges in different conductance regimes. The trace corresponds to piezo retraction that causes an initially large constriction to be pulled apart to progressively smaller sizes. The piezo elongation or retraction speed is 5 nm/s.

**FIG. 2.** (a) Size dependence of strength of silver, plotted as a function of both conductance and cross-section area of the samples. (b) Zoom-in view; inset shows more data points as well as SD of bridges less than 4 atoms in diameter. (c) Comparison of size dependence of strength for silver and gold.

**FIG. 3.** Schematic showing geometric configurations giving rise to various plateaus.

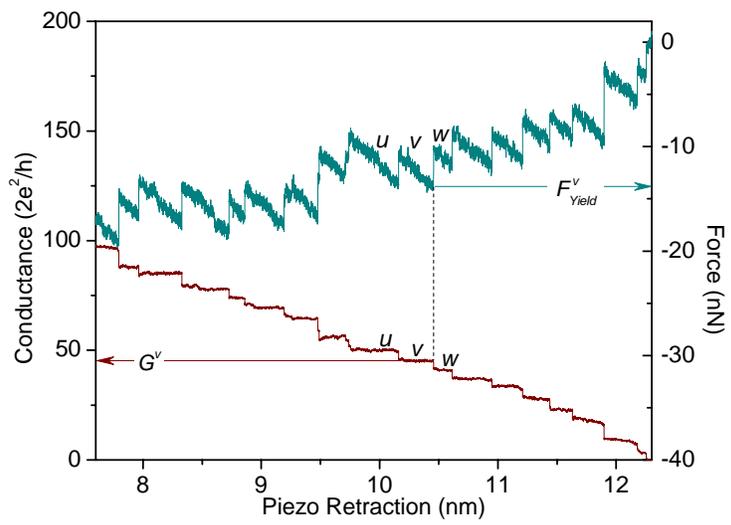

**Figure. 1**

Armstrong, Hua, Chopra, Phys Rev B-Rapids

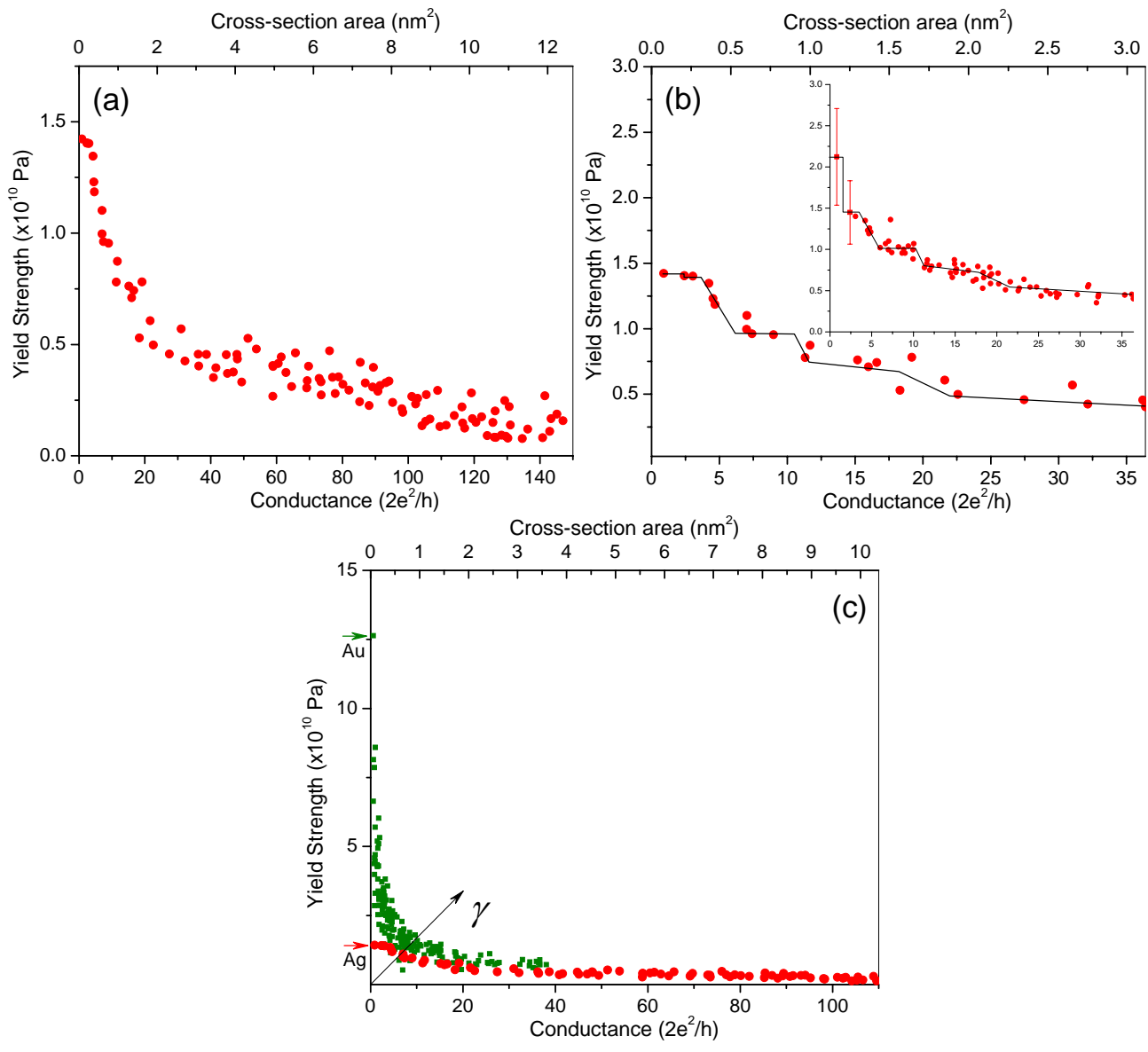

**Figure. 2**

Armstrong, Hua, Chopra, Phys Rev B-Rapids

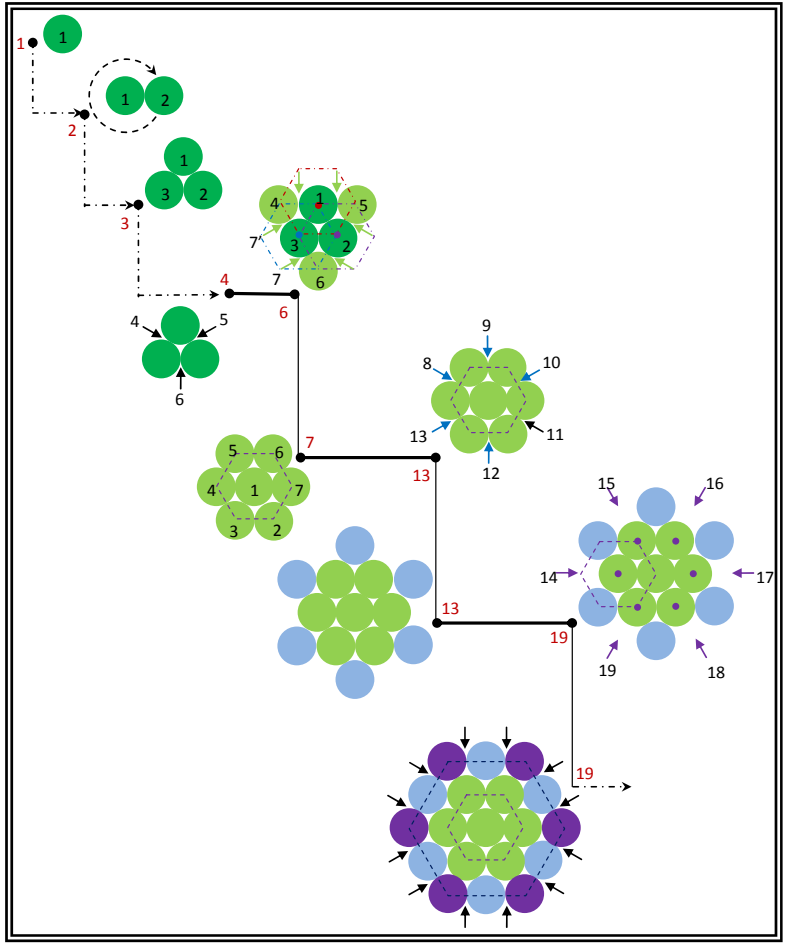

**Figure. 3**

Armstrong, Hua, Chopra, Phys Rev B-Rapids

# References

*Corresponding author. Electronic mail: hchopra@buffalo.edu